\documentclass[a4paper,12pt,reqno,superscriptaddress,nofootinbib]{revtex4}
\usepackage[centertags]{amsmath}
\usepackage{amsfonts}
\usepackage{amssymb}
\usepackage{amsthm}
\usepackage{newlfont}
\usepackage{stmaryrd}
\usepackage{mathrsfs}
\usepackage{euscript}
\usepackage{siunitx}
\usepackage{physics}
\usepackage{algorithm2e}
\usepackage{graphicx,subcaption}
\usepackage{enumitem}
\usepackage{natbib}
\usepackage{graphicx}
\usepackage{color}
\usepackage{floatrow}
\usepackage{caption}
\usepackage{hyperref}
\usepackage{hhline}
\usepackage{hyperref}
\usepackage{cleveref}
\usepackage{import}
\usepackage{comment}

\usepackage{mhchem}

\theoremstyle{plain}
  \newtheorem{theorem}{Theorem}[section]
  
  \newtheorem{proposition}[theorem]{Proposition}
  
\theoremstyle{definition}

\theoremstyle{remark}
  \newtheorem{remark}[theorem]{Remark}

\numberwithin{equation}{section}

\newcommand{\nd}{^{\mathrm{nd}}}

\newcommand{\caD}{{\mathcal D}}

\newcommand{\caF}{{\mathcal F}}

\newcommand{\caH}{{\mathcal H}}
\newcommand{\caI}{{\mathcal I}}

\newcommand{\caL}{{\mathcal L}}

\newcommand{\caS}{{\mathcal S}}

\newcommand{\opunit}{\text{1}\kern-0.22em\text{l}}

\DeclareMathAlphabet{\mathpzc}{OT1}{pzc}{m}{it}

\newcommand{\beq}{\begin{equation}}
\newcommand{\beqa}{\begin{eqnarray}}
\newcommand{\eeq}{\end{equation}}
\newcommand{\eeqa}{\end{eqnarray}}

\newcommand{\pair}[2]{(#1,#2)}

\newcommand{\id}{\textrm{d}}

\usepackage{floatrow}
\usepackage{caption}
\usepackage{color,soul}

\newcommand{\trp}{^{\mathsf T}}

\DeclareCaptionJustification{justified}{\leftskip=0pt \rightskip=0pt \parfillskip=0pt plus 1fil}
\usepackage{amsmath,amssymb,amsthm}
\usepackage{graphicx}
\usepackage{mhchem}
\usepackage{hyperref}

\theoremstyle{remark}

\begin{document}

\title{The normal approximation to macroscopic dynamical fluctuations}

\author{Christian Maes}
\affiliation{Department of Physics and Astronomy, KU Leuven, Belgium}
\author{Kate\v{r}ina Mlad\'a}
\email{mlada@fzu.cz}
\thanks{corresponding author.}
\affiliation{Institute of Physics, Czech Academy of Sciences, Prague, Czech Republic}
\author{Karel Neto\v{c}n\'{y}}
\affiliation{Institute of Physics, Czech Academy of Sciences, Prague, Czech Republic}

\begin{abstract}
Under local detailed balance, the Lagrangian of macroscopic dynamical fluctuations has a
canonical structure: its time-antisymmetric part is the entropy flux, and its time-symmetric part is a
Legendre pair of kinetic and dissipation functions. We show what survives of that structure for normal
(Gaussian) fluctuations around a nonequilibrium steady state. The quadratic Lagrangian keeps the
canonical form, with the Hessian of the dissipation function as noise matrix. However, the force on the fluctuations differs from the
linearized thermodynamic force by a term proportional to the driving and to the state dependence
of the dynamical activity, so that local detailed balance is violated at the Gaussian level. Nevertheless, the Gaussian dynamics is gradient flow
for the quadratic part of the nonequilibrium entropy, around an emergent nondissipative current.  Finally, we show which parts of the Lagrangian can be reconstructed from measured Gaussian fluctuations.
\end{abstract}


\maketitle

\section{Introduction}\label{sec:intro}

Macroscopic equations describe typical behavior; fluctuations around them carry the information about how that behavior comes about. Since Onsager~\cite{pons,O}, the linear (Gaussian) processes around equilibrium have been the main bridge between fluctuations and response: their drift is fixed by the thermodynamic force, their noise by the Onsager matrix, and time-reversal symmetry produces reciprocity and the fluctuation--dissipation theorem.\\
Away from equilibrium, the large deviations of macroscopic trajectories are known to have a canonical structure as soon as local detailed balance holds~\cite{relax}.  That is very similar to and inspired by nonlinear extensions of the relation between fluctuations and gradient flow using the (global) detailed balance condition \cite{pons,MPR13,genericmaes,Kraaij_2020}.  The time-antisymmetric part of the path-space Lagrangian can now be identified with the (irreversible) entropy flux in terms of a thermodynamic force.  Everything else is encoded in a Legendre pair of time-symmetric functions, the kinetic function $\psi$ and the dissipation function $\psi^*$. The zero-cost path gives the macroscopic relaxation equation toward nonequilibrium, and its structure extends those of gradient flows and \texttt{GENERIC}. \\

However in practice, what is most often measured or simulated are the small, normal fluctuations around the macroscopic evolution: linear-noise approximations of chemical kinetics~\cite{vanKampen,Kurtz1971}, fluctuating hydrodynamics  and Gaussian current statistics. The present paper asks what of the canonical structure survives at that level, and what the Gaussian level can tell about the full structure as an inverse problem. The main results are:
\begin{enumerate}
\item The quadratic Lagrangian of normal fluctuations around a nonequilibrium steady state keeps the canonical form (Sec.~\ref{sec:normal}). Its noise matrix $\Gamma$ is the Hessian of the dissipation function $\psi^*$ at the stationary force; its drift defines a linear force $\caF(v)$ on fluctuations.
\item That force is not the linearized thermodynamic force. The difference is  first order in the stationary driving, coming from the state dependence of the dynamical activity. Local detailed balance is thus violated at the Gaussian level (Sec.~\ref{sec:ldb}). It is restored with respect to the quadratic part of the nonequilibrium entropy, together with what we can call an emergent nondissipative current.
\item The associated Ornstein--Uhlenbeck dynamics for states and integrated currents (Sec.~\ref{sec:ou}) allows one to reconstruct the quadratic Lagrangian from data. Current noise measures the frenetic Hessian at the operating point, information that the relaxation equation alone does not contain.
\item Linear response is expressed through the Gaussian correlations (Sec.~\ref{sec:response}). Violations of the fluctuation--dissipation and Green--Kubo relations are governed by the nondissipative current.  Details are postponed to a future publication.
\end{enumerate}
Two families of examples run through the text: chemical reaction networks with one free species, including the Schl\"ogl model, and driven diffusion on a ring (Sec.~\ref{sec:examples}). Section~\ref{sec:connections} gives connections to and distinctions from previous works.  Section~\ref{sec:beyond} discusses multistability and normal fluctuations around limit cycles. Detailed computations are in the Appendices.

\section{Canonical structure of dynamical fluctuations}\label{sec:canonical}

We recall the main results of~\cite{relax,Maes_2008}. Let $z$ denote the macroscopic state (a density profile, a vector of concentrations, \dots) and $j$ a current, related by
\begin{equation}\label{eq:kinematics}
\dot z = \caD j,
\end{equation}
with $\caD$ a linear operator (minus a divergence, a stoichiometric matrix, the identity). For a system with large parameter $N$ (number of particles, volume), we assume that the paths of states and currents satisfy a large deviation principle,
\begin{equation}\label{eq:ldp}
\mathrm{Prob}\big[(z(s),j(s))_{0\le s\le t}\big]\asymp e^{-N\caI_0(z(0))}\exp\Big[-N\int_0^t\id s\,\caL\big(j(s),z(s)\big)\Big],
\end{equation}
with $\caL(j,z)\geq 0$  convex in $j$.\\

\paragraph*{Spaces and pairings.} At each state $z$ (element of a differentiable manifold) there are two pairs of dual spaces: the tangent space (velocities $\dot z$) with its dual, the cotangent space; and the space of currents $j$ with its dual, the space of forces $f$. The adjoint $\caD^\dagger$ maps cotangent vectors to forces, $\pair{a}{\caD j}=\pair{\caD^\dagger a}{j}$. In a basis, currents and forces have components $j_b$, $f_b$, where $b$ labels the transport channels: the reactions of a network, the bonds of a graph, or positions $x$ in a continuum (where sums become integrals); tangent vectors have components $a_c$, with $c$ labelling species or positions. The entropy flux is the pairing $\pair{F}{j}=\sum_b F_b\,j_b$. We set $k_B=1$, so entropies are dimensionless. Round brackets $(\cdot,\cdot)$ always denote such pairings (except when delimiting the argument of a function such as $\cal L)$; angular brackets $\langle\cdot\rangle$ denote expectations.\\

\paragraph*{Local detailed balance.} There are a thermodynamic force $F(z)$ and a Hamiltonian flow $J^H(z)$ with $\pair{F}{J^H}=0$ such that for all $\dot z = \cal D j$,
\begin{equation}\label{eq:ldb}
\caL\big(J^H(z)-j,z\big)-\caL\big(J^H(z)+j,z\big)=\pair{j}{F(z)}.
\end{equation}
\noindent {\bf Remark:}
The force may contain a gradient part $\cal D^\dagger \id S$ and a non-gradient (rotational, cycle) part $G$,
\begin{equation}\label{eq:forcesplit}
F(z)=\cal D^\dagger \id S(z)+G(z),
\end{equation}
with $S$ an equilibrium entropy; then $G\neq 0$ is the nonequilibrium driving.\\

\paragraph*{Canonical form.} The time-symmetric part of $\caL$ is free but given by a Legendre pair $(\psi,\psi^*)$ of convex functions, even in their first argument and vanishing at zero~\cite{relax}:
\begin{equation}\label{eq:Lcan}
\caL(j,z)=\psi\big(j-J^H(z),z\big)+\psi^*\Big(\frac{F(z)}{2},z\Big)-\Big(j-J^H(z),\frac{F(z)}{2}\Big) .
\end{equation}
The pair may also depend on the driving $G$; we suppress that dependence in the notation. We call $\psi^*$ the dissipation function: it determines how forces are turned into currents.\\

\paragraph*{Relaxation equation.} The zero-cost path $\caL(j_z,z)=0$ gives the typical current and the macroscopic evolution,
\begin{equation}\label{eq:relax}
j_z=J^H(z)+\partial_f\psi^*\Big(\frac{F(z)}{2},z\Big),\qquad \dot z=\caD j_z .
\end{equation}
Equilibrium corresponds to $G=0$; the entropy $S$ then increases along~\eqref{eq:relax}. Away from equilibrium there is another {\it nonequilibrium entropy} $\caS(z)$, the static fluctuation functional of the stationary state (aka quasipotential), which is a Lyapunov function~\cite{relax}.\\

\paragraph*{Assumptions.} We assume that~\eqref{eq:relax} has a stable stationary point $z^s$ (Sec.~\ref{sec:beyond} relaxes this), with stationary current $j^s=j_{z^s}$, $\caD j^s=0$, and that $\caL$ is twice continuously differentiable near $(j^s,z^s)$. For simplicity of notation we take $J^H=0$ from now on, as in both running examples; Remark~\ref{rem:JH} explains the changes for $J^H\neq 0$.

\section{Running examples}\label{sec:examples}

\subsection{Chemical networks with one free species}\label{sec:chem}

Consider a well-mixed volume $N$ with one free species $Z$ (concentration $z$) and two reactions $i=1,2$, each producing one $Z$ in its forward direction, so that $\caD=(1,\,1)$. All other species are chemostatted. With mass-action rates $r_{\pm i}(z)$,
\begin{align}
F_i(z)&=\log\frac{r_i(z)}{r_{-i}(z)},\label{eq:chemF}\\
\psi^*(f,z)&=\sum_{i=1,2}2a_i(z)\,(\cosh f_i-1),\qquad a_i=\sqrt{r_i r_{-i}},\label{eq:chempsi}
\end{align}
so that $j_i=2a_i\sinh(F_i/2)=r_i-r_{-i}$ and $\dot z=j_1+j_2$. The prefactor $a_i$ is the frenetic part of the channel; the corresponding traffic is
\begin{equation}\label{eq:activity}
t_i=r_i+r_{-i}=2a_i\cosh(F_i/2).
\end{equation}
The cycle $(1,-1)^\text{T}\in\ker\caD$ (reaction 1 forward, reaction 2 backward) has affinity $F_1-F_2$. At a stationary state, $J = j^s_1=-j^s_2$ is the stationary cycle current.\\

\paragraph*{Catalytic network.} As a first example \cite{lux5}, take
\begin{equation}\label{eq:catalytic}
\ce{A1 + Z <=>[$k_{+1}$][$k_{-1}$] B1 + 2Z},\qquad \ce{A2 <=>[$k_{+2}$][$k_{-2}$] B2 + Z},
\end{equation}
with chemostat concentrations $\alpha_{1,2},\beta_{1,2}$. In units where $k_{+1} \alpha_1=k_{+2}\alpha_2=1$, and with $\kappa_1=k_{-1}\beta_1$, $\kappa_2=k_{-2} \beta_2$,
\begin{equation}\label{eq:catrates}
r_1=z,\quad r_{-1}=\kappa_1z^2,\quad r_2=1,\quad r_{-2}=\kappa_2 z .
\end{equation}
The cycle affinity $F_1-F_2=\log(\kappa_2/\kappa_1)$ does not depend on $z$; the network is in equilibrium if and only if $\kappa_1=\kappa_2$. The stationary concentration is
\begin{equation}\label{eq:catzs}
z^s=\frac{1-\kappa_2+\Delta}{2\kappa_1},\qquad \Delta=\sqrt{(1-\kappa_2)^2+4\kappa_1},
\end{equation}
with cycle current $J=\kappa_2 z^s-1$.\\

\paragraph*{Schl\"ogl model.} The second example is Schl\"ogl's model~\cite{Schlogl1972},
\begin{equation}\label{eq:schlogl}
\ce{A + 2Z <=>[$k_1$][$k_{-1}$] 3Z},\qquad \ce{B <=>[$k_2$][$k_{-2}$] Z},
\end{equation}
with $r_1=k_1 \alpha z^2$, $r_{-1}=k_{-1}z^3$, $r_2=k_2 \beta$, $r_{-2}=k_{-2}z$. The cycle converts $\mathrm{A}\rightleftharpoons\mathrm{B}$, with affinity $\log(k_1k_{-2} \alpha /k_{-1}k_2 \beta)$. The cubic rate equation is bistable in a range of chemostat concentrations, which requires a nonzero cycle affinity (Sec.~\ref{sec:multi}).

\subsection{Driven diffusion on a ring}\label{sec:ring}

Let $\rho(x)$, $x\in[0,L)$ with periodic boundary conditions, be a density with local entropy $\oint s(\rho(x))\,\id x$ ($s$ concave), driven by a constant field $E$ along the ring. Then
\begin{align}
&\caD=-\partial_x,\qquad \caD^\dagger=\partial_x,\qquad F(\rho)=\partial_x s'(\rho)+E,\label{eq:ringF}\\
&\psi^*(f,\rho)=\oint\chi(\rho(x))\,f(x)^2\,\id x,\label{eq:ringpsi}
\end{align}
with mobility $\chi(\rho)\geq 0$. The current is $j=\chi(\rho)E-D(\rho)\,\partial_x\rho$ with diffusivity $D=-\chi s''$ (Einstein relation). For the weakly asymmetric exclusion process, $\chi=\rho(1-\rho)$, $s=-\rho\log\rho-(1-\rho)\log(1-\rho)$ and $D=1$, and the hydrodynamic equation is the viscous Burgers equation. The uniform profile $\bar\rho$ is stationary, with current $j^s=\chi(\bar\rho)E$ around the ring. The non-gradient part of the force is the field: $\oint E\,\id x=EL\neq 0$.\\
Such examples, weakly driven or boundary-driven diffusions are the main models of the macroscopic fluctuation theory developed by \cite{BDGJL2015,F,T}, as limits of stochastic interacting particle systems.\\
A nonlocal mobility (kernel) $\chi(x,y;\rho)$ and nonlocal entropy $s(\rho)$ fit the same framework. The formulas below then extend with $\chi$ read as an operator and  $s(\rho)$ as a general functional of $\rho$.  

\section{The quadratic Lagrangian}\label{sec:normal}

\subsection{Scaling and definition}

Normal fluctuations live on the scale $N^{-1/2}$. We fix a stable stationary point $(z^s,j^s)$ of \eqref{eq:relax}. Writing
\begin{equation}\label{eq:scaling}
z=z^s+\frac{v}{\sqrt N},\qquad j=j^s+\frac{\iota}{\sqrt N},
\end{equation}
the rescaled state fluctuation $v$ and current fluctuation $\iota$ satisfy $\dot v=\caD\iota$, and their path probabilities are governed by the quadratic Lagrangian
\begin{equation}\label{eq:L2def}
L_2(\iota,v)=\lim_{N\to\infty}N\,\caL\Big(j^s+\frac{\iota}{\sqrt N},\,z^s+\frac{v}{\sqrt N}\Big),
\end{equation}
in the sense that, as $N\to\infty$,
\begin{equation}\label{eq:gaussianpaths}
\mathrm{Prob}\big[(v(s),\iota(s))_{0\le s\le t}\big]\propto\exp\Big[-\int_0^t\id s\,L_2\big(\iota(s),v(s)\big)\Big].
\end{equation}
The passage from the large deviation principle~\eqref{eq:ldp} to this central-limit statement is not automatic; it holds under smoothness of the cumulant generating functions~\cite{Bryc1993}, and is a theorem for density-dependent Markov jump processes~\cite{Kurtz1971} and their linear-noise approximation~\cite{vanKampen}. We assume it throughout.

\subsection{Main formula}

Introduce the stationary force  $F^s=F(z^s)$ and three linear objects, all evaluated at the operation point $(F^s,z^s)$:
\begin{align}
\Gamma&=\partial^2_f\psi^*(F^s/2,z^s), &&\text{(noise matrix)}\label{eq:Gamma}\\
\Theta\,v&= \pair{\partial_z\partial_f\psi^*(F^s/2,z^s)}{\;v} , &&\text{(frenetic sensitivity)}\label{eq:Theta}\\
\Lambda\,v&=\frac{\id}{\id\epsilon}\,j_{z^s+\epsilon v}\Big|_{\epsilon=0}=\frac12\Gamma F'v+\Theta v, &&\text{(current response)}\label{eq:Lambda}
\end{align}
where $F'=\partial_zF(z^s)$. $\Gamma$ is symmetric and strictly positive definite (by the assumed stability) Hessian of the dissipation function on the space of forces, evaluated at steady thermodynamic force $F(z^s)/2$.  The outer round brackets in~\eqref{eq:Theta} denote the pairing of the tangent space and its dual, where linearization allows us to identify state perturbations $v$ with tangent vectors at the stationary state $z^s$. $\Lambda$ maps state fluctuations to current fluctuations. Define the force acting on state fluctuations
\begin{equation}\label{eq:calF}
\caF(v)=2\Gamma^{-1}\Lambda v=F'v+2\Gamma^{-1}\Theta v .
\end{equation}

\begin{proposition}\label{prop:L2}
The quadratic Lagrangian is
\begin{equation}\label{eq:L2}
L_2(\iota,v)=\frac12\big(\iota-\Lambda v,\;\Gamma^{-1}(\iota-\Lambda v)\big),
\end{equation}
and has the canonical form
\begin{equation}\label{eq:L2can}
L_2(\iota,v)=\psi_2(\iota)+\psi_2^*\Big(\frac{\caF(v)}{2}\Big)-\Big(\iota,\frac{\caF(v)}{2}\Big),
\end{equation}
with the quadratic Legendre pair
\begin{equation}\label{eq:psi2}
\psi_2(\iota)=\frac12\pair{\iota}{\Gamma^{-1}\iota},\qquad \psi_2^*(f)=\frac12\pair{f}{\Gamma f}.
\end{equation}
\end{proposition}

The proof (Appendix~\ref{app:L2}) uses only that $\caL(j_z,z)=0$ and $\partial_j\caL(j_z,z)=0$ for all $z$, together with $\partial_j^2\caL(j_z,z)=\partial^2_j\psi=(\partial_f^2\psi^*)^{-1}$ by Legendre duality.\\

The force $\caF(v)$ in \eqref{eq:calF} has two parts. The first, $F'v$, is the variation of the thermodynamic force. The second, $2\Gamma^{-1}\Theta v$, comes from the sensitivity of the dissipation function on the state.  Since $\psi^*$ is even in $f$, $\partial_f\psi^*(0,z)=0$ for all $z$, we have that
\begin{equation}\label{eq:Thetaeq}
F^s=0\quad\Longrightarrow\quad\Theta=0,\qquad \caF(v)=F'v .
\end{equation}
Therefore, the second part of $\caF$ is a genuinely nonequilibrium term. \\

\paragraph*{Two classes of dissipation functions.} (i) For quadratic $\psi^*(f,z)=\pair{f}{\chi(z)f}$, one has $j_z=\chi F$, $\Gamma=2\chi$ and $\caF(v)=F'v+\chi^{-1}\pair{\partial_z\chi}{v}F^s$. (ii) For channel-wise products $\psi^*(f,z)=\sum_i a_i(z)\varphi(f_i)$, as in \eqref{eq:chempsi}, with $\varphi$ even and convex,
\begin{equation}\label{eq:calFprod}
\caF_i(v)=F'_iv+2\,\pair{\partial_z\log a_i}{\,v}\,\frac{\varphi'(F^s_i/2)}{\varphi''(F^s_i/2)},
\end{equation}
which for $\varphi(f)=2(\cosh f-1)$ gives the factor $\tanh(F_i/2)=j_i/t_i$, the ratio of current to traffic \cite{Maes2008Steady,maes2007entropy}.

\subsection{Examples}

\paragraph*{One-species networks.} See Example \ref{sec:chem}.  Here $\Gamma=\mathrm{diag}(t_1,t_2)$ and $\Lambda=(\lambda_1,\lambda_2)\trp$ with $\lambda_i=r'_i-r'_{-i}$ (derivatives at $z^s$). Then $\caF_i=2\lambda_i/t_i$ per unit $v$, and by~\eqref{eq:calFprod},
\begin{equation}\label{eq:chemcalF}
\caF_i=F'_i+2\tanh\frac{F^s_i}{2}\,(\log a_i)'=F'_i+\Big(\frac{r'_i}{r_i}+\frac{r'_{-i}}{r_{-i}}\Big)\frac{j_i}{t_i}.
\end{equation}
For the catalytic network, $F'_1=F'_2=-1/z^s$, $(\log a_1)'=3/(2z^s)$ and $(\log a_2)'=1/(2z^s)$, so that
\begin{equation}\label{eq:catcalF}
\caF_1=\frac{-1+3J/t_1}{z^s},\qquad \caF_2=\frac{-1-J/t_2}{z^s}
\end{equation}
in terms of the cycle current $J$.\\

\paragraph*{Ring.} See Example \ref{sec:ring}.  At the uniform state, $\Gamma=2\chi(\bar\rho)$ and
\begin{equation}\label{eq:ringLambda}
\Lambda v=c\,v-D\,\partial_xv,\qquad c=\chi'(\bar\rho)\,E,
\end{equation}
with $D=-\chi s''$ at $\bar\rho$, and
\begin{equation}\label{eq:ringcalF}
\caF(v)=s''(\bar\rho)\,\partial_xv+\frac{c}{\chi(\bar\rho)}\,v .
\end{equation}
The first term is $F'v$; the second is the ``correction.'' For exclusion, $c=E(1-2\bar\rho)$ is the drift velocity in the linearized dynamics.\\

\begin{remark}\label{rem:JH}
With a Hamiltonian flow $J^H(z)$, $\Lambda$ and consequently $\caF(v)$ remain unchanged, and both formulas~\eqref{eq:L2} as well as~\eqref{eq:L2can} hold with $\iota$ replaced by $\iota-\pair{\partial_zJ^H}{v}$.
\end{remark}

\section{Local detailed balance at the Gaussian level}\label{sec:ldb}

\subsection{Formal structure and its mismatch}

Reversing the current fluctuation at fixed $v$ in~\eqref{eq:L2can} gives
\begin{equation}\label{eq:ldb2}
L_2(-\iota,v)-L_2(\iota,v)=\pair{\iota}{\caF(v)} .
\end{equation}
The quadratic Lagrangian therefore has the form of local detailed balance, with $\caF(v)$ in the role of the force. Local detailed balance proper would require that to be the entropy flux of the fluctuation. However, by~\eqref{eq:calF},
\begin{equation}\label{eq:mismatch}
\caF(v)-F'v=2\Gamma^{-1}\Theta v ,
\end{equation}
which vanishes only at equilibrium. It also vanishes if the dissipation function does not depend on the state at the operating point, $\Theta=0$.  In conclusion, the antisymmetric part of $L_2$ need not be the physical entropy flux.\\
The origin of the mismatch is the time reversal itself. Physical time reversal maps $j^s+\iota/\sqrt N$ to $-j^s-\iota/\sqrt N$. When $j^s\neq 0$, the reversed path lies {\it outside} the Gaussian window: $\caL(-j^s,z^s) - \caL(j^s,z^s)=\pair{j^s}{F(z^s)}>0$ is the stationary entropy production, and its probability is exponentially small in $N$. What remains at the Gaussian level is the reversal $\iota\mapsto-\iota$ around the stationary current, which is not a symmetry of the physics. The antisymmetric part~\eqref{eq:ldb2} then also picks up how the current responds kinetically to the state, through $\Theta$.

\subsection{Gaussian entropy and nondissipative current}

Local detailed balance can be restored at the Gaussian level, but with respect to new objects. Let $C = \langle v \; v^{\mathsf T}\rangle$ be the stationary covariance of $v$ (Sec.~\ref{sec:ou}), which solves the Lyapunov equation (for derivation see Appendix~\ref{app:ou})
\begin{equation}\label{eq:lyap}
KC+CK^\dagger+B=0,\qquad K=\caD\Lambda,\quad B=\caD\Gamma\caD^\dagger .
\end{equation}
Define the Gaussian entropy $\caS_2(v)=-\frac12\pair{v}{C^{-1}v}$ and
\begin{equation}\label{eq:F0Jnd}
\caF_0(v)=\caD^\dagger\nabla\caS_2(v)=-\caD^\dagger C^{-1}v,\qquad J\nd(v)=\Lambda v-\frac12\Gamma\caF_0(v).
\end{equation}

\begin{proposition}\label{prop:emergent}
(a) $L_2(\iota,v)=\psi_2\big(\iota-J\nd(v)\big)+\psi_2^*\big(\caF_0(v)/2\big)-\pair{\iota-J\nd(v)}{\caF_0(v)/2}$, so that
\begin{equation}\label{eq:ldbemergent}
L_2\big(J\nd(v)-\iota,v\big)-L_2\big(J\nd(v)+\iota,v\big)=\pair{\iota}{\caF_0(v)}.
\end{equation}
(b) $J\nd$ is nondissipative: $\pair{J\nd(v)}{\caF_0(v)}=0$ for all $v$.\\
(c) $\caD J\nd(v)=-QC^{-1}v$ with $Q= -KC-\frac12B$ antisymmetric. In particular $\caS_2$ is conserved by the flow $\caD J\nd$.
\end{proposition}

The proof is in Appendix~\ref{app:emergent}. The Gaussian dynamics is thus the gradient flow of $\caS_2$ with mobility $\Gamma/2$, plus a nondissipative current $J\nd$, exactly the structure~\eqref{eq:Lcan} with a Hamiltonian-like part. In other words, the zero-cost flow for $L_2$ is part of the {\tt GENERIC} family \cite{Kraaij_2020,genericmaes}.  However, from the validity of the normal approximation we must have $C^{-1}=-\caS''(z^s)$ (shown in Proposition~\ref{prop:Sdd}, when $\caS$ is twice differentiable at $z^s$). 
Hence, $\caS_2$ (defined under \eqref{eq:lyap})  is the quadratic part of the nonequilibrium entropy $\caS$ of~\cite{relax}, {\it not} of the equilibrium entropy $S$. Proof is in Appendix~\ref{app:Sdd}.\\

Comparing with physical local detailed balance, two violations appear, both vanishing at equilibrium:
\begin{enumerate}
\item the Gaussian force $\caF_0(v)$ differs from $F'v$ whenever the stationary covariance differs from the equilibrium one;
\item the emergent current $J\nd$ is nonzero. Its part outside $\ker\caD$ moves the state while conserving $\caS_2$ (a circulation $Q\neq 0$); its part in $\ker\caD$ is a cycle current driven by the state fluctuation.
\end{enumerate}
At equilibrium $G=0$, $J^H=0$: $K=\frac12BS''$, $C^{-1}=-S''$, $\caF_0(v)=F'v$ and $J\nd=0$ and the standard Onsager (linear) processes are recovered. 

\subsection{Examples}

\paragraph*{One-species networks.} See Example \ref{sec:chem}. The state is one-dimensional, so $Q=0$, and we can solve \eqref{eq:lyap},
\begin{equation}\label{eq:chemC}
K=\lambda_1+\lambda_2=\frac12(t_1\caF_1+t_2\caF_2),\qquad C=-\frac{t_1+t_2}{2K}.
\end{equation}
See also the end of Appendix~\ref{app:Sdd}. The emergent current is a pure cycle current,
\begin{equation}\label{eq:omega}
J\nd(v)=\omega\,v\begin{pmatrix}1\\-1\end{pmatrix},\qquad \omega=\frac{\lambda_1t_2-\lambda_2t_1}{t_1+t_2}=\frac{t_1t_2}{2(t_1+t_2)}(\caF_1-\caF_2).
\end{equation}
The physical force has no fluctuating cycle component: $F_1-F_2$ is fixed by the chemostats, so $F'_1=F'_2$. The Gaussian force does: $\caF_1-\caF_2$ is, by~\eqref{eq:chemcalF}, proportional to the stationary cycle current. For the catalytic network,
\begin{equation}\label{eq:catomega}
\caF_1-\caF_2=\frac{J}{z^s}\Big(\frac{3}{t_1}+\frac1{t_2}\Big),\qquad \omega=\frac{J\,(t_1+3t_2)}{4z^s(1+z^s)},
\end{equation}
and, using $t_1+t_2=2(1+z^s)$ and $K=-\Delta$,
\begin{equation}\label{eq:catFano}
C=\frac{1+z^s}{\Delta},\qquad \frac{C}{z^s}=\frac{1+z^s}{z^s\Delta}.
\end{equation}
At equilibrium ($\kappa_1=\kappa_2=\kappa$, $z^s=1/\kappa$, $\Delta=1+\kappa$) the Fano factor (variance of the number of molecules divided by its mean) is $C/z^s$ and equals one, which is the Poisson value you would predict from the equilibrium expression of the entropy $S(z) = - z \log(z/z^\text{eq}) + z$.\\

\paragraph*{Ring.} See Example \ref{sec:ring}.  On mean-zero fluctuations, $K=D\partial_x^2-c\,\partial_x$ and $B=-2\chi\partial_x^2$, and the Lyapunov equation is solved by the local-equilibrium covariance $C=-1/s''(\bar\rho)$ times the identity (Appendix~\ref{app:ring}). Hence $\caF_0(v)=F'v$: the Gaussian entropy is the local-equilibrium one, as for the product invariant measure of exclusion. The ``generalized local detailed balance''  \eqref{eq:ldbemergent} holds with current,
\begin{equation}\label{eq:ringJnd}
J\nd(v)=c\,v,\qquad Q=-\frac{c}{s''(\bar\rho)}\,\partial_x ,
\end{equation}
which transports fluctuations at the drift velocity $c=\chi'(\bar\rho)E$ without changing $\caS_2$. For exclusion at half filling, $c=0$: there, normal fluctuations cannot be told apart from equilibrium ones.

\subsection{Intuitive meaning}

At equilibrium a fluctuation relaxes because the thermodynamic force pulls it back, and its reversal costs exactly the entropy it produces. Around a driven steady state, a fluctuation also changes how active the transport channels are, and the stationary current through those channels then changes even at fixed force. Seen from the fluctuation, that kinetic effect looks like an extra force, which is the term $2\Gamma^{-1}\Theta v$. It is proportional both to how strongly the activity depends on the state and to how hard the system is driven. In the chemical example, a fluctuation in $z$ changes the activity of the two reactions unequally and so drives a cycle current, even though the cycle affinity itself is fixed. On the ring, a density fluctuation changes the mobility and so the field-driven current, and the fluctuation is carried along. The Gaussian description absorbs all such effects into its own entropy and its own nondissipative current.

\section{Normal dynamics}\label{sec:ou}
We come to the second point of discussion about the normal approximation, the Gaussian dynamics itself.
\subsection{Langevin equations}

The Gaussian path measure~\eqref{eq:gaussianpaths} with Lagrangian~\eqref{eq:L2} is that of the linear stochastic dynamics
\begin{equation}\label{eq:langevin}
\iota(t)=\Lambda v(t)+\eta(t),\qquad \dot v(t)=\caD\iota(t),\qquad \langle\eta(t)\eta(t')^{\mathsf T}\rangle=\Gamma\,\delta(t-t'),
\end{equation}
with white noise $\eta$ on the space of currents. For the state alone,
\begin{equation}\label{eq:OUv}
\dot v=Kv+\xi,\qquad K=\caD\Lambda,\qquad \langle\xi(t)\xi(t')^{\mathsf T}\rangle = B\,\delta(t-t'),\qquad B=\caD\Gamma\caD^\dagger .
\end{equation}
The integrated current fluctuation $I(t)=\int_0^t\iota(s)\,\id s$ satisfies $\caD I(t)=v(t)-v(0)$ and
\begin{equation}\label{eq:OUI}
\dot I=K_II+\Lambda v(0)+\eta,\qquad K_I=\Lambda\caD,
\end{equation}
with the intertwining $\caD K_I=K\caD$. The noise matrix $\Gamma$ is the same for all channels whose currents are measured: the Hessian has become the strength of the current noise in \eqref{eq:langevin}.

When $\caD$ has a nontrivial co-kernel (conservation laws, as for the ring), $K$ has a kernel and the fluctuations are restricted to the subspace compatible with the conservation laws, on which $K$ is assumed stable. All inverses below are taken on that subspace.

\subsection{Stationary correlations}

The stationary covariance $C$ solves~\eqref{eq:lyap}, and $\langle v(t)v(0)^{\mathsf T}\rangle=e^{Kt}\,C$ for $t\geq0$ (Onsager regression). For current fluctuations, for $t>0$,
\begin{equation}\label{eq:currcorr}
\langle\iota(t)\iota(0)^{\mathsf T}\rangle=\Lambda e^{Kt}\big(C\Lambda^\dagger+\caD\Gamma\big),
\end{equation}
plus the white-noise part $\Gamma\delta(t)$. The integrated current has a diverging variance, linear in time, along $\ker\caD$.

\begin{proposition}\label{prop:sigma}
The long-time covariance rate of the integrated current is
\begin{equation}\label{eq:sigma}
\Sigma=\lim_{t\to\infty}\frac1t\,\langle I(t)I(t)^{\mathsf T}\rangle=\Pi\,\Gamma\,\Pi^\dagger,\qquad \Pi=1-\Lambda K^{-1}\caD .
\end{equation}
$\Pi$ is the (oblique) projector onto the cycle space $\ker\caD$ along the image space $\mathrm{im}\,\Lambda$ of state-induced currents.  The limit $e^{tK_I}\to\Pi$ as $t\to\infty$ for $K_I$ in \eqref{eq:OUI}. Equivalently, $\Sigma=\int_{-\infty}^\infty\langle\iota(t)\iota(0)\trp\rangle\,\id t$.
\end{proposition}
The proof is in Appendix~\ref{app:ou}. The long-time current fluctuations are therefore the current noise projected onto cycles, along the directions in which currents follow the state.\\

Note: At equilibrium, the image space $\mathrm{im}\,\Lambda$ is the $\Gamma^{-1}$-orthogonal complement of $\ker\caD$, so that $\Pi$ is the orthogonal projector in the noise metric $\pair{x}{\Gamma^{-1}y}$. Out of equilibrium, $\Theta$ tilts the image space $\mathrm{im}\,\Lambda$ and $\Pi$ becomes oblique. This obliqueness is what breaks the Green--Kubo relation by adding a frenetic contribution; more on that in Sec.~\ref{sec:response}.\\

\paragraph*{One-species networks.} See Example \ref{sec:chem}. $\ker\caD$ is spanned by $(1,-1)$. Writing $I=I_c\,(1,-1)+\dots$, the cycle current has variance rate
\begin{equation}\label{eq:Dc}
\Sigma_c=\frac{\lambda_2^2t_1+\lambda_1^2t_2}{K^2}=\frac{t_1t_2\,(t_1\caF_1^2+t_2\caF_2^2)}{(t_1\caF_1+t_2\caF_2)^2}.
\end{equation}
At equilibrium, $\caF_1=\caF_2$ and $\Sigma_c=t_1t_2/(t_1+t_2)$: the two reactions act as resistors in series.\\

\paragraph*{Ring.} See Example \ref{sec:ring}. The spatially averaged current $\bar I=L^{-1}\oint I\,\id x$ obeys $\dot{\bar I}=\bar\eta$, because $\oint\Lambda v\,\id x=c\oint v\,\id x=0$ by mass conservation. Hence $\Sigma_{\bar I}=2\chi(\bar\rho)/L$, unchanged by the drive at the Gaussian level.

\subsection{Gaussian reconstruction}\label{sec:reconstruct}

The quadratic Lagrangian is fully identified from stationary time series of states and integrated currents.  Starting from the Langevin equations \eqref{eq:langevin} and \eqref{eq:OUI} we find:
\begin{equation}\label{eq:identify}
\Gamma=\lim_{\Delta t\to0}\frac{\langle\Delta I\,\Delta I^{\mathsf T}\rangle}{\Delta t},\qquad \Lambda=\lim_{\Delta t\to0}\frac{\langle\Delta I\,v(t)^{\mathsf T}\rangle}{\Delta t}\;C^{-1},
\end{equation}
with $\Delta I=I(t+\Delta t)-I(t)$; that is, the noise matrix is the quadratic variation of the currents, and $\Lambda$ is the regression of current increments on the state. From states alone, one obtains only $K=\caD\Lambda$, as $\frac{\id}{\id t}\langle v(t)v(0)^{\mathsf T}\rangle|_{0^+}C^{-1}$, and $B=\caD\Gamma\caD^\dagger$ from~\eqref{eq:lyap}: the cycle components of $\Gamma$ and $\Lambda$ remain invisible. From currents alone, short times give $\Gamma$, long times give $\Sigma$, and the regression of current correlations gives $K_I$, as in~\eqref{eq:OUI}.\\

That identifies part of the original Lagrangian. The noise matrix is the Hessian $\partial_f^2\psi^*$ at the stationary force, while the relaxation equation only gives the gradient $\partial_f\psi^*$ at that force. If $F'$ is known from thermodynamics, $\Theta=\Lambda-\frac12\Gamma F'$ gives the mixed derivative $\partial_z\partial_f\psi^*$.\\ For chemical networks that is enough to recover everything locally: the mean current gives $r_i-r_{-i}$ and the current noise gives $t_i=r_i+r_{-i}$, hence both rates, hence $a_i$, $F_i$ and $\caL$ at $z^s$. Varying a control parameter, such as a chemostat concentration, moves the stationary state; repeating the measurement along the way reconstructs the rate functions, and hence $\caL$, over the range of states that can be made stationary, provided the dependence of the rates on the control parameter is known, as it is for mass-action kinetics.\\
 In general, if the driving can be varied at fixed state, integrating $\frac{\id^2}{\id s^2}\psi^*(sF/2,z)=\frac14\pair{F}{\Gamma(s)F}$ along the ray recovers $\psi^*(\cdot,z)$ there.\\

\section{Connections with earlier Gaussian theories and fluctuating hydrodynamics}\label{sec:connections}

The algebra of the Gaussian level is standard: a linear stochastic dynamics, its
stationary covariance from a Lyapunov equation, and a split of the drift into a
symmetric and an antisymmetric part.  What the canonical structure adds is the extension of the Langevin dynamics by including the current variables, and the physical identification of all objects: the current noise as the Hessian of the dissipation function at the operating point, and the linear force derived from (the Hessian of) the nonequilibrium entropy in combination with an emergent current from the state dependence of the dynamical activity. \\
In the following, we compare our results with previous studies and make the connections explicit.

\subsection{Gaussian fluctuations around steady states}

Lax~\cite{Lax1960} treated Gaussian fluctuations around nonequilibrium steady states as
linear processes whose stationary covariance $C$ solves a Lyapunov equation, our
\eqref{eq:lyap}, and stressed that $C$ is in general not the inverse Hessian of the
equilibrium entropy. Keizer~\cite{Keizer1987} built on this a statistical
thermodynamics of nonequilibrium processes. Two of his ingredients reappear here.
First, in his canonical theory of elementary processes the noise intensity of each
process is fixed by the sum of its forward and backward rates. That is exactly our
noise matrix for a channel-wise dissipation function: for
$\psi^*(f,z)=\sum_i 2a_i(z)(\cosh f_i-1)$, the Hessian at the operating force is
$\Gamma_{ii}=2a_i\cosh(F_i/2)=t_i=r_i+r_{-i}$, the traffic. Keizer's postulate is thus
the statement that the noise matrix is $\partial_f^2\psi^*$ at the stationary force,
not at zero force. Second, Keizer proposed a nonequilibrium entropy whose second
derivative at the steady state is $-C^{-1}$, and suggested measuring it through
fluctuations. Proposition~\ref{prop:Sdd} 
derives that identification from large
deviations: $-C^{-1}$ is the Hessian of the nonequilibrium entropy $\mathcal S$, the
quasipotential of the stationary distribution. Keizer's entropy is therefore our
Gaussian entropy $\caS_2$, and it is the quadratic part of a globally defined object.\\

Of course, knowing $\caS_2$ does not determine the
Gaussian dynamics: by Proposition~\ref{prop:emergent} the drift is the gradient flow of
$\caS_2$ plus the emergent current $J^\text{nd}$, and it is $J^\text{nd}$ that will control the
violations of fluctuation--dissipation and Green--Kubo relations. Moreover,
working with currents rather than states, we keep track of cycle components of $\Gamma$ 
and $J^\text{nd}$ that a description in terms of states alone, as in~\cite{Lax1960,Keizer1987},
cannot see. In the one-species networks the state is one-dimensional and the entire
nonequilibrium signature at the Gaussian level is such a cycle current.

\subsection{Nonequilibrium potentials}

For a diffusion $\dot z=b(z)+\sqrt{2/N}\,\sigma(z)\xi$ with diffusion matrix
$B=\sigma\sigma^T$, Graham~\cite{Graham1977} and Graham and T\'el~\cite{GrahamTel1984,GrahamTel1986}
defined the nonequilibrium potential $\Phi$ as the quasipotential of the stationary
distribution, $P^\text{st}\asymp e^{-N\Phi}$. It solves the Hamilton--Jacobi equation
$(\nabla\Phi,b)+(\nabla\Phi,B\nabla\Phi)=0$, which is equivalent to the decomposition
\begin{equation}
b=-B\nabla\Phi+r,\qquad (\nabla\Phi,\,r)=0 ,
\label{eq:grahamtel}
\end{equation}
of the drift into a part driven by the potential and a remainder orthogonal to its
gradient. The remainder conserves $\Phi$, and the time-reversed (adjoint) dynamics has
drift $-B\nabla\Phi-r$.\\

Proposition~\ref{prop:emergent}(c) is the Gaussian instance of \eqref{eq:grahamtel}.
With $\Phi_2=-\caS_2=\frac12(v,C^{-1}v)$ and $B=\mathcal D\Gamma\mathcal D^\dagger$, the
state drift splits as
\begin{equation}
Kv=-\tfrac12 B\,C^{-1}v+\mathcal D J^\text{nd}(v),\qquad
\mathcal D J^\text{nd}(v)=-QC^{-1}v,\qquad Q^\dagger=-Q,
\end{equation}
and $\mathcal DJ^\text{nd}$ is orthogonal to $\nabla\Phi_2=C^{-1}v$.\\
Furthermore, here, the
decomposition is lifted from states to currents, Eq.~\eqref{eq:ldbemergent}, where it
takes the form of a generalized local detailed balance with respect to $\caS_2$.\\

Kwon, Ao and Thouless~\cite{KwonAoThouless2005} showed that any linear stochastic
dynamics $\dot x=-Fx+\zeta$, $\langle\zeta\zeta^T\rangle=2D\,\delta$, can be rewritten
exactly as
\begin{equation}
(S+T)\,\dot x=-U x+\xi,\qquad S=S^T\ge0,\quad T=-T^T,\quad U=U^T,
\label{eq:KAT}
\end{equation}
with $\langle\xi\xi^T\rangle=2S\,\delta$ and stationary distribution
$\propto e^{-\frac12(x,Ux)}$. The construction solves the Lyapunov equation:
$U=C^{-1}$, and $(S+T)^{-1}=D+Q$ with $Q$ antisymmetric, so that $F=(D+Q)\,U$.
In our notation, $D=\frac12B$ and $F=-K$, and the identity is our
\begin{equation}
K=-\big(\tfrac12B+Q\big)\,C^{-1},\qquad Q=-KC-\tfrac12 B=-Q^\dagger ,
\end{equation}
the same algebra as Proposition~\ref{prop:emergent}(c). The decomposition
\eqref{eq:KAT} is thus a mathematical fact about linear processes, and our
Proposition~\ref{prop:emergent} is in that sense its restatement.\\

The physics enters
through three identifications that \cite{KwonAoThouless2005} does not make: (i) the
potential $U=C^{-1}$ is the Hessian of the nonequilibrium entropy of the underlying
nonlinear dynamics (Proposition~\ref{prop:Sdd}); (ii) the symmetric part $D$ is
$\frac12\mathcal D\Gamma\mathcal D^\dagger$, the projection of the Hessian at
the operating point; and (iii) the antisymmetric part $Q$ has a definite origin,
the frenetic sensitivity $\Theta$ in the force \eqref{eq:calF}, so that it is of first
order in the stationary driving and proportional to the state dependence of the
activity. In particular, $Q\neq0$ is a measurable signature of driving with
state-dependent kinetics, not an arbitrary gauge choice.

\subsection{Fluctuating hydrodynamics and macroscopic fluctuation theory}

Fluctuating hydrodynamics adds noise to the hydrodynamic equations, following Landau and
Lifshitz~\cite{LandauLifshitzFM}: random stress and heat-flux terms are added to the
Navier--Stokes equations, with covariance fixed by fluctuation--dissipation. This theory was extended and rigorously proven  by \cite{BDGJL2015,F,T} for large class of weakly driven or boundary-driven diffusions as limits of stochastic interacting particle systems.\\
In our framework, fluctuating hydrodynamics is the special case of a quadratic
dissipation function, $\psi^*(f,\rho)=\int\chi(\rho)f^2\,\id x$, which is exactly
Example~\ref{sec:ring}, Eq.~\eqref{eq:ringpsi}. The quadratic Lagrangian $L_2$ is then
linearized fluctuating hydrodynamics, with $\Gamma=2\chi(\bar\rho)$, and macroscopic fluctuation theory (MFT) is the
large-deviation level of the same theory. MFT contains the decomposition that our
Gaussian results linearize. With $V$ the quasipotential, the hydrodynamic current splits
as
\begin{equation}
J(\rho)=-\chi(\rho)\,\partial_x\frac{\delta V}{\delta\rho}+A(\rho),\qquad
\int\partial_x\frac{\delta V}{\delta\rho}\;A(\rho)\,\id x=0 ,
\label{eq:MFTsplit}
\end{equation}
which is equivalent to the Hamilton--Jacobi equation for $V$, and the adjoint
hydrodynamics, which governs the optimal fluctuation paths, reverses the first part and
keeps $A$~\cite{BDGJL2015}. Linearizing \eqref{eq:MFTsplit} around $\bar\rho$ gives
Proposition~\ref{prop:emergent}: $V$ becomes $-\caS_2$, the force
$-\partial_x\delta V/\delta\rho$ becomes $\caF_0$, and $A$ becomes $J^\text{nd}$. For the
ring, $A$ linearizes to the transport current $J^\text{nd}=c\,v$ with
$c=\chi'(\bar\rho)E$, and the Gaussian entropy is the local-equilibrium one, in
agreement with the product invariant measure of the weakly asymmetric exclusion process
on a ring.\\

The comparison also shows what is specific to the present setting.
In fluctuating hydrodynamics and MFT the noise is fixed by the near-equilibrium mobility. Our general result, that the noise matrix is the Hessian $\partial_f^2\psi^*$ at the operating force, has no counterpart there.

\section{Outlook}\label{sec:beyond}

\subsection{Multistability: the Schl\"ogl model}\label{sec:multi}

When the relaxation equation has several stable stationary points, as the Schl\"ogl model~\eqref{eq:schlogl} does for a range of parameters, the normal approximation applies around each of them separately, on times short compared with the switching time, which grows exponentially in $N$ with the barrier in the nonequilibrium entropy $\caS$. All formulas of Secs.~\ref{sec:normal}--\ref{sec:ou} for one-species networks apply at each stable point, with
\begin{equation}\label{eq:schloglK}
\lambda_1=2k_1 \alpha z^s-3k_{-1}(z^s)^2,\qquad \lambda_2=-k_{-2}.
\end{equation}
Bistability requires a nonzero cycle affinity, since with detailed balance the rate equation has a unique stable point~\cite{NicolisPrigogine1977}. Consequently $\caF_1\neq\caF_2$ at each of the two stable points, and both carry an emergent cycle current $\omega v$. As a stable point approaches a saddle-node bifurcation, $K=\lambda_1+\lambda_2\to0$, so $C$ and $\Sigma_c$ diverge as $1/|K|$ and $1/K^2$. The normal approximation then breaks down: fluctuations become non-Gaussian, on a larger scale in $N$.

\subsection{Limit cycles}\label{sec:cycles}

Write the relaxation equation~\eqref{eq:relax} as $\dot z=X(z)$, with $X(z)=\caD j_z$, and suppose it has a stable periodic orbit $\gamma(t)$ of period $T$, as in the Brusselator~\cite{PrigogineLefever1968}. Limit cycles need at least two state variables, so they do not occur for one-species networks. Writing
\begin{equation}
z=\gamma(t)+\frac{v}{\sqrt N},\qquad j=j_{\gamma(t)}+\frac{\iota}{\sqrt N},
\end{equation}
the derivation of Appendix~\ref{app:L2}, with the stationary point replaced by the point $\gamma(t)$ on the orbit, gives a time-periodic quadratic Lagrangian
\begin{equation}\label{eq:L2cycle}
L_2(\iota,v;t)=\frac12\big(\iota-\Lambda(t)v,\;\Gamma(t)^{-1}(\iota-\Lambda(t)v)\big),
\end{equation}
with $\Gamma(t)=\partial_f^2\psi^*\big(F(\gamma(t))/2,\gamma(t)\big)$ and $\Lambda(t)v=\frac{\id}{\id\epsilon}j_{\gamma(t)+\epsilon v}\big|_{\epsilon=0}$. The associated Gaussian dynamics is that of \eqref{eq:langevin}, now with periodic coefficients:
\begin{equation}\label{eq:OUcycle}
\iota=\Lambda(t)v+\eta,\qquad \dot v=K(t)v+\caD\eta,\qquad \langle\eta(t)\eta(t')^T\rangle=\Gamma(t)\,\delta(t-t'),
\end{equation}
with $K(t)=\caD\Lambda(t)$. The state noise $\caD\eta$ has covariance $B(t)=\caD\Gamma(t)\caD^\dagger$, as in \eqref{eq:OUv}.\\
It is possible to record the time along the limit cycle, called the asymptotic phase $\phi$: each $z$ in the basin of attraction of the orbit has an asymptotic phase $\phi(z)\in\mathbb R/T\mathbb Z$: the solution starting at $z$ converges to $\gamma(t+\phi(z))$. Hence $\phi(\gamma(s))=s$ and $\phi(z(t))=\phi(z(0))+t$ along solutions. The level sets of $\phi$ are the isochrons.  To first order, $\phi\big(\gamma(t)+v/\sqrt N\big)=t+\theta/\sqrt N$ defining the phase fluctuation $\theta$.  It turns out that $\theta$ accumulates noise without any restoring force: it is a time-changed Brownian motion.  For further details, we refer to~\cite{Gaspard2002}.
 Bounds on the number of coherent oscillations in terms of the cycle affinities have been conjectured~\cite{BaratoSeifert2017}.\\

Local detailed balance is violated here in an essential way. On a limit cycle, the reversed orbit $\gamma(-t)$ does not solve the relaxation equation, and by \eqref{eq:ldb} its cost per period is $N\sigma_T$ with
\begin{equation}
\sigma_T=\oint\big(j_{\gamma(t)},F(\gamma(t))\big)\,\id t>0,
\end{equation}
the entropy produced per period. Time reversal thus maps the Gaussian window around $\gamma$ into a window around an exponentially improbable path, and no reversal symmetry, not even a modified one, acts on the phase. 

\subsection{Linear response}\label{sec:response}
In Section~\ref{sec:ou} all expectations $\langle\cdot\rangle$ are taken in the Gaussian dynamics~\eqref{eq:langevin}, that is, in the normal approximation. The variables $v$ and $\iota$ are the rescaled fluctuations of~\eqref{eq:scaling}, their stationary law is ${\cal N}(0,C)$ with $C$ from~\eqref{eq:lyap}, and the time correlations are $\langle v(t)v(s)^\text{T}\rangle=e^{K(t-s)}C$ and~\eqref{eq:currcorr}. They are the limits as $N\uparrow \infty$ of $N$ times the connected (cumulant) correlations of the state and current of the original process, at fixed times. Higher connected correlations (third, fourth,...) are smaller still, of order $N^{-2}$ and beyond. At leading order only the covariance survives, so the process is Gaussian. The correlations are those within a single well of a stable stationary point.\\

At the macroscopic level the linear response is the same as the Gaussian one. Consider a perturbation $F\to F+h$ \emph{at fixed dissipation function}, i.e., $\psi^*$ unchanged. The typical current becomes $j^h_z=\partial_f\psi^*\big(\frac12(F(z)+h),z\big)$, and expanding around $z^s$ to first order in $\delta z=z-z^s$ and $h$,
\begin{equation}
j^h_z=j^s+\Lambda\,\delta z+\tfrac12\Gamma h,\qquad
\delta\dot z=K\,\delta z+\tfrac12\caD\Gamma h,
\label{eq:maclin}
\end{equation}
where the $h$-term is the Hessian $\Gamma=\partial_f^2\psi^*(F^s/2,z^s)$ acting on $h/2$. On the other hand, in the Gaussian dynamics, $\iota=\Lambda v+\frac12\Gamma h+\eta$ and
$\langle\dot v\rangle=K\langle v\rangle+\frac12\caD\Gamma h$. The latter coincides with \eqref{eq:maclin} upon identifying $\delta z=\langle v\rangle/\sqrt N$ and taking the perturbation of order $N^{-1/2}$ at the Gaussian level, $h\to h/\sqrt N$, so that the shifted state remains in the Gaussian window. Hence, starting from the steady state,
\begin{equation}
\delta z(t)=\int_{-\infty}^t e^{K(t-s)}\,\tfrac12\caD\Gamma h(s)\,\id s,\qquad
\delta j(t)=\Lambda\,\delta z(t)+\tfrac12\Gamma h(t),
\end{equation}
and for constant $h$ the stationarity condition $\caD\delta j=0$ shifts the state by $\delta\bar v=-\frac12K^{-1}\caD\Gamma h$, giving $\delta j=\frac12\Pi\Gamma h$.\\

The results on linear response will be part of  subsequent work.

\subsection{Ballistic regime}

Our scaling, with drift and noise both of order $N$ in the action as for
density-dependent jump processes~\cite{Kurtz1971}, is the diffusive or reactive one.
For strongly asymmetric lattice gases and for integrable systems the relevant scaling is
Euler (ballistic), $x\sim t\sim L$. The hydrodynamics is then hyperbolic, for instance
the inviscid Burgers equation $\partial_t\rho+\partial_x[\rho(1-\rho)]=0$ for the totally
asymmetric exclusion process, or generalized hydrodynamics for integrable models, with
dissipation only at shocks.\\

Fluctuation theories exist at that scale, but none with the canonical structure of
Sec.~\ref{sec:canonical}. In ballistic macroscopic fluctuation theory~\cite{DoyonBMFT2023} there is no
dynamical noise at leading order: fluctuations come from the randomness of the initial
state and are transported deterministically by the Euler flow, and the action is the
initial-state large-deviation functional pushed along characteristics. For the totally
asymmetric exclusion process, the large deviations of Jensen and Varadhan at Euler
scale~\cite{Jensen2000,Varadhan2004} have a rate function given by the entropy
production of non-entropic weak solutions, so that the only time-antisymmetric cost sits
at shocks. Beyond the Gaussian window, asymmetric systems show
Kardar--Parisi--Zhang fluctuations with $t^{1/3}$ scaling~\cite{DerridaLebowitz1998},
so a Gaussian theory can only describe short times or weak asymmetry; on the ring this
is the window in which our drift velocity $c$ transports fluctuations without distorting
them.\\

In the language of Sec.~\ref{sec:canonical}, the ballistic limit is the one in which the Hamiltonian
current $J^H$, the Euler current that conserves the entropy of smooth solutions, is of
order one, while the force $F$ and the noise $\Gamma$ are suppressed by $1/L$. The
canonical structure then degenerates: the zero-cost flow is purely Hamiltonian, the
dynamical Legendre pair $(\psi,\psi^*)$ disappears at leading order, and local detailed
balance reduces to the time-reversal invariance of the Euler flow, supplemented by
entropy production at shocks. There is therefore no canonical formalism for dynamical
fluctuations at the ballistic scale in the present sense. Ballistic MFT takes its
place, and the diffusive corrections, at Navier--Stokes order, are where $F$ and
$\Gamma$ re-enter as weak dissipation on top of an Euler $J^H$. That ``weakly
dissipative'' setting, with $J^H$ the Euler current and the canonical pair at order
$1/L$, is a natural extension of the present work.

\section{Conclusion}\label{sec:conclusion}

Most often measured or simulated around a stable nonequilibrium steady state are the normal fluctuations.  In this paper we asked 1) what survives of local detailed balance and of the canonical structure of macroscopic dynamical fluctuations, 2) whether and how the normal fluctuations and perhaps the larger dynamical fluctuations as well, can be reconstructed.\\

The quadratic Lagrangian of the fluctuations $(v,\iota)$ keeps the canonical form with a quadratic dissipation
function and a linear force $\caF(v)$ acting on the
state fluctuation. The noise matrix is the Hessian of $\psi^*$ at the actual stationary force
$F^\text{s}/2$. $\caF$ is not the linearized thermodynamic force
$F'v$: it contains an additional term expressing the sensitivity of the dynamical activity
to the state, proportional to the stationary driving. It
vanishes when the dissipation function does not depend on the state. As a result, normal fluctuations around steady
nonequilibrium are Onsager processes with respect to a modified entropy plus a nondissipative flow, that is of {\tt GENERIC}-type.\\
Secondly, the Gaussian path measure is that of a linear (Ornstein--Uhlenbeck) dynamics, for the state and for the integrated current.  The quadratic Lagrangian can be identified
from stationary time series. \\

We mention that for linear response, the Gaussian description reproduces the macroscopic response, and the emergent nondissipative
current is expected to control the violation of fluctuation--dissipation and Green--Kubo relations.  

\section*{Acknowledgements}
KM is supported by Charles University, project GA UK No. 414225.

\bibliographystyle{unsrt}
\bibliography{allpapers,newrefs}

\newpage
\appendix

\section{Derivation of the quadratic Lagrangian $L_2$}\label{app:L2}

Write $\caL(j,z)$ in the form~\eqref{eq:Lcan} with $J^H=0$. For every $z$, $j_z$ minimizes $\caL(\cdot,z)$ with $\caL(j_z,z)=0$, so $\partial_j\caL(j_z,z)=0$. By Legendre duality, $\partial_j\psi(j_z,z)=F(z)/2$ and
\begin{equation}
\partial_j^2\caL(j_z,z)=\partial^2_j\psi(j_z,z)=\big[\partial^2_f\psi^*(F(z)/2,z)\big]^{-1}.
\end{equation}
Taylor's theorem in $j$ at fixed $z$ gives $\mathcal{L}(j_z+\delta,z)=\frac12 (\delta, \; \partial_j^2\mathcal{L}(j_z,z)\,\delta)+O(|\delta|^3)$. With~\eqref{eq:scaling}, $j_z=j^s+\Lambda v/\sqrt N+O(1/N)$, so $\delta=(\iota-\Lambda v)/\sqrt N+O(1/N)$, and $\partial_j^2\mathcal{L}(j_z,z)=\Gamma^{-1}+O(N^{-1/2})$. The derivative of $j_z=\partial_f\psi^*(F(z)/2,z)$ in the direction $v$ is $\Lambda v = \frac12\Gamma F'v+\Theta v$, which is~\eqref{eq:Lambda}. Multiplying by $N$ and letting $N\to\infty$ gives~\eqref{eq:L2}.  Expanding~\eqref{eq:L2} and using $\mathcal{F}(v)/2=\Gamma^{-1}\Lambda v$ gives~\eqref{eq:L2can}. 

\section{Gaussian process}\label{app:ou}

From~\eqref{eq:OUv}, $v(t)=e^{Kt}v(0)+\int_0^te^{K(t-s)}\caD\eta(s)\,\id s$, which gives $\langle v(t)v(0)^{\mathsf T}\rangle=e^{Kt}C$.  Taking the equal-time covariance and imposing stationarity yields the Lyapunov equation~\eqref{eq:lyap} (see, e.g., \cite{gardiner}). For $t>0$, $\langle\eta(0)v(t)^{\mathsf T}\rangle=\Gamma\caD^\dagger e^{K^\dagger t}$, which gives~\eqref{eq:currcorr}.

\paragraph*{Proof of Proposition~\ref{prop:sigma}.} Integrating~\eqref{eq:OUv} gives $\int_0^tv\,\id s=K^{-1}\big(v(t)-v(0)\big)-K^{-1}\caD\int_0^t\eta\,\id s$. Hence
\begin{equation}
I(t)=\Lambda K^{-1}\big(v(t)-v(0)\big)+\Pi\int_0^t\eta(s)\,\id s .
\end{equation}
This can be read as a decomposition to the space of cycle currents and state-induced currents. The first term is bounded in time, so $\Sigma=\Pi\Gamma\Pi^\dagger$. \\
Properties of $\Pi$: a) $\caD\Pi=\caD-KK^{-1}\caD=0$, so $\mathrm{im}\,\Pi\subset\ker\caD$, b) $\Pi x=x$ for $x\in\ker\caD$, c) $\Pi\Lambda=\Lambda-\Lambda K^{-1}K=0$. Thus $\Pi$ is the projector onto $\ker\caD$ along $\mathrm{im}\,\Lambda$. Moreover $e^{tK_I}=1+\Lambda\big(\int_0^te^{sK}\id s\big)\caD\to1-\Lambda K^{-1}\caD$. \\
The equivalence with $\int\langle\iota(t)\iota(0)^{\mathsf T}\rangle\id t=\Gamma-\Lambda K^{-1}(C\Lambda^\dagger+\caD\Gamma)-(\Lambda C+\Gamma\caD^\dagger)K^{-\dagger}\Lambda^\dagger$ follows from expanding $\Pi\Gamma\Pi^\dagger$ and using $K^{-1}C+CK^{-\dagger}=-K^{-1}BK^{-\dagger}$, which is~\eqref{eq:lyap}.

\section{Proof of Proposition~\ref{prop:emergent}}\label{app:emergent}

Since $\Lambda v=J\nd(v)+\frac12\Gamma\caF_0(v)$, expanding~\eqref{eq:L2} around $J\nd$ gives (a) exactly as in Appendix~\ref{app:L2}. For (c), define $Q=-KC-\frac12B$. The Lyapunov equation~\eqref{eq:lyap} gives $Q+Q^\dagger=-(KC+CK^\dagger)-B=0$. Then
\begin{equation}
\caD J\nd(v)=Kv+\tfrac12\caD\Gamma\caD^\dagger C^{-1}v=\big(K+\tfrac12BC^{-1}\big)v=-QC^{-1}v .
\end{equation}
For (b), $\pair{J\nd}{\caF_0}=\pair{\caD J\nd}{\nabla\caS_2}=\pair{QC^{-1}v}{C^{-1}v}=0$ by antisymmetry of $Q$. A part of $J\nd$ in $\ker\caD$ is orthogonal to $\caF_0\in\mathrm{im}\,\caD^\dagger$ automatically. \\
At equilibrium, $\Lambda=\frac12\Gamma\caD^\dagger S''$ and $C^{-1}=-S''$ solve~\eqref{eq:lyap}, so $J\nd=0$.

\section{The Hessian of the nonequilibrium entropy}\label{app:Sdd}

\paragraph*{Hamiltonian on force space.} For a force $f$ (an element of the dual of the space of currents at $z$) define
\begin{equation}\label{eq:Hdef}
\caH(z,f)=\sup_j\big\{\pair{f}{j}-\caL(j,z)\big\},
\end{equation}
the Legendre transform of $\caL$ in $j$. It is the cumulant generating function of the current at fixed state. With the canonical form~\eqref{eq:Lcan}, the substitution $k=j-J^H(z)$ and Legendre duality $\psi^{**}=\psi^*$ give
\begin{equation}\label{eq:Hcan}
\caH(z,f)=\pair{f}{J^H(z)}+\psi^*\Big(\tfrac12F(z)+f,\,z\Big)-\psi^*\Big(\tfrac12F(z),\,z\Big).
\end{equation}
For $J^H=0$, local detailed balance~\eqref{eq:ldb} is equivalent to the symmetry
\begin{equation}\label{eq:Hsym}
\caH(z,f)=\caH\big(z,-F(z)-f\big),
\end{equation}
since $\caL(-j,z)=\caL(j,z)+\pair{j}{F(z)}$ turns $\sup_j\{\pair{-F-f}{j}-\caL(j,z)\}$ into $\caH(z,f)$ after $j\to-j$. The same follows directly from~\eqref{eq:Hcan} and the evenness of $\psi^*$.\\

\paragraph*{Hamilton--Jacobi equation.} The quasipotential $V=-\caS$ is the rate function of the stationary measure, $P^\text{st} (z)\asymp e^{-NV(z)}$,
\begin{equation}
V(z)=\inf\Big\{\int_{-\infty}^0\caL(j,z)\,\id s:\ \dot z=\caD j,\ z(-\infty)=z^s,\ z(0)=z\Big\}.
\end{equation}
Wherever $V$ is differentiable,
\begin{equation}\label{eq:HJ}
\caH\big(z,\caD^\dagger \id V(z)\big)=0,\qquad V(z^s)=0 .
\end{equation}

\begin{proposition}\label{prop:Sdd}
Let $V=-\caS$ be $C^2$ near $z^s$, solve~\eqref{eq:HJ}, and have a strict local minimum at $z^s$, $M:=V''(z^s)>0$. Let $K=\caD\Lambda$ be stable and $B=\caD\Gamma\caD^\dagger$ be positive definite (both on the subspace compatible with the conservation laws, if any). Then
\begin{equation}
M=C^{-1},\qquad\text{that is}\qquad C^{-1}=-\caS''(z^s),
\end{equation}
where $C$ is the unique solution of the Lyapunov equation~\eqref{eq:lyap}.
\end{proposition}

\begin{proof}
Put $z=z^s+x$. Since $V(z^s)=0$ and $z^s$ is a minimum, $\partial V(z)=Mx+O(|x|^2)$, hence $f=\caD^\dagger\partial V=\caD^\dagger Mx+O(|x|^2)$. Expanding~\eqref{eq:Hcan} in $f$ at fixed $z$, using that $\partial_f\psi^*(F/2,z)+J^H(z)=j_z$ and $\partial^2_f\psi^*(F/2,z)=\Gamma(z)$,
\begin{equation}
\caH(z,f)=\pair{f}{j_z}+\tfrac12\pair{f}{\Gamma(z)f}+O(|f|^3).
\end{equation}
Now $\caD j^s=0$ and $j_z=j^s+\Lambda x+O(|x|^2)$, so
\begin{equation}
\pair{\caD^\dagger Mx}{j_z}=\pair{Mx}{\caD j_z}=\pair{Mx}{Kx}+O(|x|^3).
\end{equation}
Inserting $f=\caD^\dagger Mx$ gives, to second order in $x$,
\begin{equation}
0=\caH\big(z,\caD^\dagger\partial V\big)=\pair{Mx}{Kx}+\tfrac12\pair{Mx}{B\,Mx}+O(|x|^3).
\end{equation}
A quadratic form vanishes for all $x$ if and only if its symmetric part does. Using $M=M^\dagger$ we get
\begin{equation}\label{eq:Riccati}
MK+K^\dagger M+MBM=0 .
\end{equation}
Since $M>0$ is invertible, multiplying~\eqref{eq:Riccati} from the left and from the right by $M^{-1}$ gives $KM^{-1}+M^{-1}K^\dagger+B=0$, i.e.\ $M^{-1}$ solves~\eqref{eq:lyap}. For stable $K$ the Lyapunov equation has a unique solution, so $M^{-1}=C$.
\end{proof}

For Example~\ref{sec:chem}, $\caD=(1,1)$, $V'(z)=\log\big(w_-(z)/w_+(z)\big)$ with $w_\pm$ the sums of the forward and reverse rates, so $V''(z^s)=-K/w$ with $w=w_+(z^s)=w_-(z^s)=\frac12(t_1+t_2)$. That equals $C^{-1}$ of~\eqref{eq:chemC}. For the Schl\"ogl example with $(k_1\alpha,\,k_{-1},\,k_{-2},\,k_2\beta)=(6,1,11,6)$,
the rate equation is $\dot z=-(z-1)(z-2)(z-3)$, with stable points $z=1$ and $z=3$.
At $z=1$: $K=-2$, $w=12$, and $V''(1)=1/6=C^{-1}$.

\section{Driven diffusion on a ring}\label{app:ring}

In Fourier modes $e^{ikx}$, $k=2\pi n/L$, $n\neq0$: $K(k)=-Dk^2-ick$ and $B(k)=2\chi k^2$, all at $\bar\rho$. The Lyapunov equation $2\,\mathrm{Re}K(k)\,C(k)+B(k)=0$ gives $C(k)=\chi/D=-1/s''$, independent of $k$. Thus $\caF_0(v)=\partial_x(s''v)=F'v$, and $J\nd(v)=\Lambda v+\chi\,\partial_x(v/C)=cv-D\partial_xv+D\partial_xv=cv$. Each Fourier mode decays at rate $Dk^2$ and rotates at frequency $ck$; the rotation is the circulation $Q$.

\end{document}